\documentclass[12pt]{article}
\usepackage{amsmath}
\usepackage{times}
\usepackage{graphicx}
\usepackage{color}
\usepackage{multirow}
\usepackage{amssymb}
\usepackage{booktabs}

\usepackage{rotating}
\usepackage{latexsym}
\usepackage{subfigure} 
\usepackage{amsthm}
\newtheorem{definition}{Definition}
\newtheorem{proposition}{Proposition}

\newcommand{\captionfonts}{\normalsize}

\makeatletter
\long\def\@makecaption#1#2{%
  \vskip\abovecaptionskip
  \sbox\@tempboxa{{\captionfonts #1: #2}}%
  \ifdim \wd\@tempboxa >\hsize
    {\captionfonts #1: #2\par}
  \else
    \hbox to\hsize{\hfil\box\@tempboxa\hfil}%
  \fi
  \vskip\belowcaptionskip}
\makeatother

\begin{document}
\hspace{13.9cm}1

\ \vspace{20mm}\\


{\LARGE Non-uniqueness phenomenon of object representation in modelling IT cortex by deep convolutional neural network (DCNN)}

\ \\
{\bf  Qiulei Dong$^{\displaystyle 1, \displaystyle 2, \displaystyle 3}$, Bo Liu$^{\displaystyle 1, \displaystyle 2}$, Zhanyi Hu$^{\displaystyle 1, \displaystyle 2, \displaystyle 3, *}$}\\
{$^{\displaystyle 1}$National Laboratory of Pattern Recognition, Institute of Automation, Chinese Academy of Sciences, Beijing
100190, China.}\\
{$^{\displaystyle 2}$University of Chinese Academy of Sciences, Beijing 100049, China.}\\
{$^{\displaystyle 3}$CAS Center for Excellence in Brain Science and Intelligence Technology, Beijing 100190, China.}\\
{$^*$ Corresponding author: huzy@nlpr.ia.ac.cn}\\
%


\thispagestyle{empty}
\markboth{}{NC instructions}
\ \vspace{-0mm}\\

%
\section*{Abstract}
Recently DCNN (Deep Convolutional Neural Network) has been advocated as a general and promising modelling approach for neural object representation in primate inferotemporal cortex. In this work, we show that some inherent non-uniqueness problem exists in the
DCNN-based modelling of image object representations. This non-uniqueness phenomenon reveals to some extent the theoretical limitation of this general modelling approach, and invites due attention to be taken in practice.

\section*{Author summary}
In the field of neuroscience, DCNN  has been advocated recently as a general and promising modelling approach for neural object representation in primate inferotemporal cortex. However, the following uniqueness problem on the fundamental premise of this modelling approach is still unclear: does there exist a unique representation in the penultimate layer of a DCNN for a given set of image stimuli by only optimizing the object categorization performance? This problem has a great influence on the theoretical foundation and generality of the DCNN-based modelling approach. In this work, we provided a theoretical analysis on this problem as well as some supporting experimental results, and showed that there exists a non-uniqueness phenomenon of object representation under the DCNN-based modelling approach. Hence, we suggest that when DCNNs are used for modeling sensory cortex as a general framework, it is necessary for people to be aware of this potential and inherent non-uniqueness problem, and appropriate network architectures in DCNN learning should be carefully considered.


\section{Introduction} \label{intro}

Object recognition is a fundamental task of a biological vision system. It is widely believed that the primate  inferotemporal (IT) cortex   is the final neural site for visual object representation. Due to viewpoint change, illumination variation and other factors, how visual objects are represented in IT cortex, which manifests sufficient invariance to such identity-orthogonal factors, is still largely an open issue in neuroscience.

There are many different natural and manmade object categories, and each category in turn contains various different members. Neuroscientists generally believed that ``the computational goal of object representation is likely the same across all of IT cortex'' \cite{Chang2017}, although special cortical areas do exist for face, body parts, buildings, etc. Currently, a number of works in neuroscience advocate the DCNN (Deep Convolutional Neural Network) as a new framework for modelling vision and brain information processing \cite{Khaligh2014,Cadieu2014}. In \cite{Yamins2014,Yamins2016}, DCNN is regarded as a promising general modelling approach for understanding sensory cortex, called ``the goal-driven approach''.

The basic idea of the goal-driven approach for IT cortex modelling can be summarized as: \textit{a multi-layered DCNN is trained by ONLY optimizing the object categorization performance with a large set of visual category-labeled objects. Once a high categorization performance is achieved, the outputs of the penultimate layer neurons of the trained DCNN, which are regarded as the object representation, can reliably predict the IT neuron spikes for other visual stimuli in rapid object recognition.\footnote{The goal-driven approach is for modelling IT neuron representation in rapid object vision, which is assumed largely a feed forward process, hence could be modelled by DCNNs which are also feed forward networks.}.}
 In addition, the outputs of the upstream layer neurons can also predict the V4 neuron spikes. The goal-driven approach is conceptually eloquent and has been successfully used to model IT cortex in rapid object recognition and predict category-orthogonal properties \cite{Hong2016}.

\section{Does the goal-driven approach satisfy the uniqueness requirement  in modelling IT cortex?}

\subsection{Motivation} \label{motivation}
Although some experimental results have demonstrated the success of the goal-driven approach in modelling IT
cortex to some extent as mentioned above, the following uniqueness problem   on \textbf{the fundamental premise of the goal-driven approach} is still unclear:  \textbf{does there exist a unique
pattern of activations of the neurons (units) in the penultimate layer of a DCNN to a given set of image stimuli by only optimizing the object categorization performance?}
This uniqueness problem on object representation via a DCNN has a great influence on the theoretical foundation and generality of the goal-driven approach in particular, and the DCNN as a new framework for vision modelling in general.

In this work, we aim to provide a theoretical analysis on this problem as well as some supporting experimental results. In order to analyse this problem more clearly, we firstly introduce the definition of DCNN layer's object representation as used for predicting the neuron responses of primate IT cortex in the aforementioned goal-driven approach:
\begin{definition} \label{def1}
For a layer of a DCNN for object recognition, the activations of the neurons in this layer to an input object image is defined as its object representation.
\end{definition}

Following the convention in the computational neuroscience, the following  representation equivalence is introduced to evaluate whether the object representations learnt from two DCNNs are the same or not:
\begin{definition} \label{def2}
Given a set of object image stimuli, if the two object representations of two DCNNs on these stimuli can be related by a linear transformation, they are considered equivalent, or the same representations.  Otherwise, they are different representations.
\end{definition}

In the deep learning community, a recent active research topic is called ``convergent learning'' \cite{Li2016}, referring whether different DCNNs can learn the same representation at the level of neurons or groups of neurons. A generally reached conclusion is that different DCNNs with the same network architecture but trained only with different random initializations, have largely different representations at the level of neurons or groups of neurons, although their image categorization performances are similar.
Note that although Li et al.'s work and the goal-driven approach focus on the representation from different points of view, the representations in the two works are closely related. Hence, the results in \cite{Li2016} could also re-highlight the aforementioned uniqueness problem in object representation via a DCNN to some extent.

Addressing this uniqueness problem, we show in the following section that, in theory, by only optimizing the image categorization accuracy, different DCNNs can give different object representations  though they have exactly the same categorization accuracy. In other words, the obtained object representations by DCNNs under the goal-driven approach could be inherently  non-unique, at least in theory.

\subsection{Theoretical analysis and experimental results}

\begin{proposition} \label{p1}
If the `Softmax' function is used as the final classifier for image categorization in modelling $N$ categories of objects via a DCNN, and the object category with the largest probability is chosen as the final categorization, and if $x=(x_1, x_2, \cdots, x_N)^T\in R^N$ is the final output of this DCNN for an input image object $I$, $f(\cdot)$ is a univariate nonlinear  monotonically increasing function, $y \triangleq  (y_1, y_2, \cdots, y_N)^T = F(x) = (f(x_1), f(x_2), \cdots, f(x_N))^T$, then $x$ and $y$ give exactly the same categorization result.
\end{proposition}


\textbf{Proof:} For $x$ and $y$, their corresponding probability vectors by Softmax are respectively:
\begin{align}
C_x = \left( \frac{e^{x_1}}{\sum_{i=1}^N e^{x_i}}, \frac{e^{x_2}}{\sum_{i=1}^N e^{x_i}}, \cdots, \frac{e^{x_N}}{\sum_{i=1}^N e^{x_i}} \right)^T  \\
C_y = \left( \frac{e^{y_1}}{\sum_{i=1}^N e^{y_i}}, \frac{e^{y_2}}{\sum_{i=1}^N e^{y_i}}, \cdots, \frac{e^{y_N}}{\sum_{i=1}^N e^{y_i}} \right)^T
\end{align}
Since $y_i = f(x_i)$ ($i=1,2, \cdots, N$) and $f(\cdot)$ is a monotonically increasing function, the magnitude order of elements for $x$  and $y$ does not change. Then the magnitude order of the two probability vectors $C_x$ and $C_y$ does not change. Since the object category with the largest probability is chosen as the final categorization, both the indices of the largest elements in $C_x$ and $C_y$  are the same, hence the same categorization results are obtained for $x$ and $y$.   \qquad \qquad \qquad \qquad  \qquad \qquad \qquad \qquad \qquad \qquad \qquad \qquad
$\blacksquare$

\textbf{Remark 1:} Since $f(\cdot)$ is a nonlinear function, $x$ and $y$ cannot be related by a linear transformation. In addition, in the deep learning community, the Softmax function is commonly used to convert the output vector of the network into a probability vector, and the category with the largest probability value is chosen as the final category.

\textbf{Remark 2:} In theory, $f(\cdot)$ could be different for different input image $I$. More generally, even the demand of monotonicity for $f(\cdot)$  is unnecessary, we need only the index of the largest value in $y$ is the same to that in $x$ because only the largest value determines the correct categorization. For the Top-$K$ categorization accuracy, we need the index set of the $K$ largest values in $y$ keep the same to that in $x$, and the rest elements are not required. Hereinafter, for the notational convenience in discussion and practicality of implementation, we always assume $f(\cdot)$ is a univariate nonlinear monotonically increasing function.

\begin{figure}[t]
\centering
  \subfigure {  \includegraphics[width=14 cm]{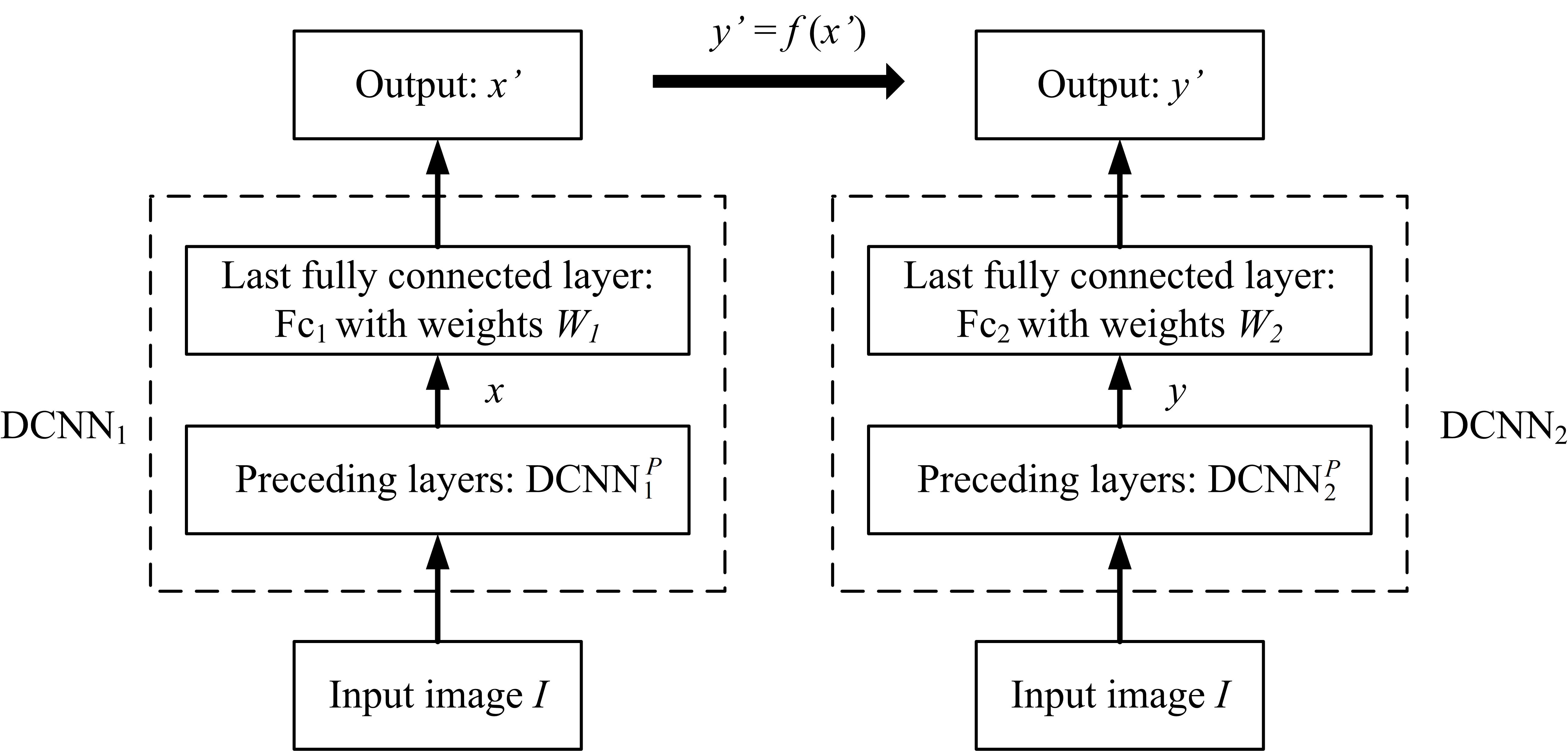}  }
\caption{DCNN$_1$ and DCNN$_2$ give the different object representations $x$ and  $y$ for the same input image object $I$, however their object categorization performances are exactly the same if $y'=f(x')$, where $f(\cdot)$ is an element-wise  nonlinear monotonically increasing function.} \label{D12}
\end{figure}

\begin{proposition} \label{p2}

As shown in Figure \ref{D12}, assume that DCNN$_{1}$ is a multi-layered network, concatenating a sub-network DCNN${}^P_{1}$ whose output is $x$, and a fully connected layer with weight matrix $W_1 \in R^{N\times M}$ and bias $b_1 \in R^{N\times 1}$ ($\{M, N\}$ are the numbers of neurons at the penultimate layer and last layer of DCNN$_\mathit{1}$ respectively, with $M > N$), with $x'=W_1 x + b_1$. And assume that DCNN$_{2}$ is a multi-layered network, concatenating a sub-network DCNN${}^P_{2}$ whose output is $y$, and a fully connected layer with weight matrix $W_2 \in R^{N\times M}$ and bias $b_2\in R^{N\times 1}$, with $y'=W_2 y + b_2$. If $y'=f(x')$ in element-wise mapping where $f(\cdot)$ is a monotonically increasing function, then the object representation $x$ under DCNN$_\mathit{1}$ cannot be related by a linear transformation to the object representation $y$ under DCNN$_{2}$, or $x$ and $y$ are two different object representations under the goal-driven approach.
\end{proposition}

\textbf{Proof:} Since  $y'=f(x')$ in element-wise mapping where $f(\cdot)$ is a monotonically increasing function, according to Proposition \ref{p1}, DCNN$_1$ and DCNN$_2$ have the identical image object categorization performance.

Since $x'=W_1 x + b_1$, then $x=(W_1^T W_1)^{+} W_1^T (x' - b_1)$, where $A^{+}$ denotes the pseudo-inverse of matrix $A$. Similarly, $y=(W_2^T W_2)^{+} W_2^T (y' - b_2)$. By Proposition 1, $x'$ and $y'$ is related by a nonlinear function, then $x$ and $y$ cannot be related by a linear transformation either. In other words, $x$ and $y$ are two different object representations under the goal-driven approach. \qquad \qquad \qquad \qquad \qquad \qquad \qquad \qquad \qquad \qquad \qquad  $\blacksquare$

\textbf{Remark 3:} Since $\{W_1, W_2\} \in R^{N\times M}$ and $M > N$ in Proposition \ref{p2}, the pseudo-inverse operator is used in the above proof. Here are a few words on the pseudo-inverse: Since $M > N$, which is the usual case in most existing DCNNs for object categorization \cite{Krizhevsky2012,Simonyan2014,Szegedy2015}, the inverse $(W_i^T W_i )^{+}$($i=1,2$) is not unique , but the equalities in $x=(W_1^T W_1 )^{+} W_1^T (x' - b_1)$ and $y=(W_2^T W_2 )^{+} W_2^T (y' - b_2)$ can be strictly met.

Proposition \ref{p2} indicates that given DCNN$_1$ with output $x'$, if there exists another multi-layered network DCNN$_2$ to output $y' = f(x')$, their representations  $x$ and $y$ would be different but with identical categorization performance. This means that  the aforementioned non-uniqueness problem in object representation modelling under the goal-driven approach  would arise regardless of how many training images are used, and how many exemplar images in each category are included. In other words, the non-uniqueness problem is an inherent problem in DCNN modelling under the goal-driven approach, and it cannot be completely removed by using more training data, at least in theory.

In the above, an implicitly assumption is that given a DCNN$_1$ with the output $x_i'$, there always exists a DCNN$_2$ with the output $y_i' = f(x_i')$.
Does such a DCNN$_2$ really always exist? This issue can be separately addressed for the following two cases. The first one is that DCNN$_1$ and DCNN$_2$ could be of different architectures, and the second one is that they are of the same architecture, but merely initialized differently during training.

\noindent \textbf{The different architecture case}

\begin{proposition} \label{p3}
There always exists a multi-layered network to map $I_i$ to $y_i$ for the given input-output pairs $\{(I_i \leftrightarrow y_i ),i=1,2, \cdots,n\}$ in Proposition \ref{p2}.
\end{proposition}

\textbf{Proof:}
As shown in Proposition \ref{p2} and Figure \ref{D12}, since DCNN$_1$ exists, it maps $I$ to $x$. Denote this mapping function as $x = S_1(I) = DCNN^P_\mathit{1}(I)$. Since $x '=W_1 x + b_1$, $y'=F(x')=((f(x_1'), f(x_2'), \cdots, f(x_n'))$, $y'=W_2 y + b_2$, and  $y=(W_2^T W_2 )^{+} W_2^T (y' - b_2)$, we have:
\begin{align}
y=(W_2^T W_2 )^{+} W_2^T (y'-b_2) = (W_2^T W_2 )^{+} W_2^T (F(W_1 S_1(I) + b_1) - b_2) \label{eq3}
\end{align}
This is just the required mapping function. By the Universal Approximation Theorem in \cite{Balazs2001}, there always exists a DCNN, denoted as DCNN$_2$, whose sub-network DCNN${}^P_2$ is able to approximate this function. \qquad \qquad \qquad \qquad \qquad \qquad \qquad \qquad $\blacksquare$

Proposition \ref{p3} indicates that given a DCNN$_1$, there always exists a DCNN$_2$ whose architecture may be different from DCNN$_1$, so that the object representations of the two DCNNs are different but with the same categorization performance. 
A training procedure  is described in the Appendix, to show how to train such a pair of DCNN$_1$ and DCNN$_2$.

\textbf{Remark 4:}
In the proof, the only requirement for DCNN$_2$ is that it should have sufficient capacity to represent the input object set, but it does not necessarily have a similar network architecture to DCNN$_1$. Note that the sufficient representational capacity is an implicit necessary requirement for any DCNN-based applications.

\textbf{Remark 5:}
In the proof, the number of input images is assumed to be unknown. However for the finite-input case, Theorem 1 in \cite{Tian2017} guarantees that there exists a two-layered neural network with ReLU activation and ($2n + d$) weights, which could represent any mapping function from input to output on sample of size $n$ in $d$ dimensions. Of course, such a constructed network could be of a memorized neural network, i.e., it can ensure the given finite inputs to be mapped to the required outputs, but it cannot guarantee that the constructed network could possess sufficient generalization ability for new samples.

\noindent \textbf{The same architecture case}

When DCNN$_1$ and DCNN$_2$ are obtained with the same network architecture but only trained under different random initializations, clearly a theoretical proof is impossible. However, based on the reported results in the ``convergent learning'' literatures as well as our  simulated experimental results, it seems they still largely have non-equivalent object representations although they have similar categorization performances.

\textbf{(1) Non-uniqueness results from ``convergent learning'' literatures}

Using AlexNet \cite{Krizhevsky2012} as a benchmark, Li et al. \cite{Li2016} showed that by keeping the architecture unchanged but only trained with different random initializations, the obtained 4 DCNNs have similar categorization performances, but their object representations are largely different in terms of one-to-one, one-to-many, and many-to-many linear representation mapping. Note that the many-to-many mapping in \cite{Li2016} is closely related to the equivalence  representation in Definition \ref{def2}. Hence, the 4 representations are largely non-equivalent and this non-equivalence becomes more prevalent with increasing convolutional layers.

	 By introducing the concepts of ``$\epsilon$-simple match set'' and ``$\epsilon$-maximum match set'', Wang et al. \cite{WangNIPS2018} showed that for the 2 representative DCNNs, VGG \cite{Simonyan2014} and ResNet \cite{He2016}, the size of maximum match set between the activation vectors of individual neurons at the same layer of the two DCNNs, which are also obtained with only different initializations as did in \cite{Li2016}, is tiny compared with the number of the neurons at that layer. It was further found that only the outputs of neurons in the $\epsilon$-maximum match set can be approximated within $\epsilon$-error  bound by a linear transformation, which indicates that for majority of the neurons at the same layer, their outputs cannot be reasonably approximated by a linear transformation, or the corresponding object representations are largely not equivalent.
	
\textbf{(2) Non-uniqueness results from our experiments}

\begin{definition}
If two DCNNs, DCNN$_1$ and DCNN$_2$,  have similar image categorization performances with  the same network architecture but different parameter configurations, they are called the similar performing pair of DCNNs.
\end{definition}

\renewcommand\arraystretch{0.6}
\begin{table}[t]
  \footnotesize
  \centering
  \begin{tabular}{|c|c|c|c|c|c|} \hline
    \multicolumn{6}{|c|}{ConvNet Configuration} \\
    \hline
    D1 & D2 & D3 & D4 & D5 & D6 \\
    \hline
    5 Layers & 8 Layers & 8 Layers & 8 Layers & 15 Layers & 9 Layers \\
    \hline
    \multicolumn{6}{|c|}{Input(32*32 RGB Image)} \\
    \hline
    Conv5-32 & Conv3-bn-32 & Conv3-bn-64 & Conv3-bn-128 & Conv3-bn-32 & Conv3-bn-64 \\
    & Conv3-bn-32 & Conv3-bn-64 & Conv3-bn-128 & Conv3-bn-32 &  \\
    &  &  &  & Conv3-bn-32 &  \\
    &  &  &  & Conv3-bn-32 &  \\
    \hline
    \multicolumn{6}{|c|}{Max-pool} \\
    \hline
    Conv5-32 & Conv3-bn-64 & Conv3-bn-128 & Conv3-bn-256 & Conv3-bn-64 & Conv3-bn-128 \\
      & Conv3-bn-64 & Conv3-bn-128 & Conv3-bn-256 & Conv3-bn-64 &  \\
      &  &  &  & Conv3-bn-64 &  \\
      &  &  &  & Conv3-bn-64 &  \\
    \hline
    \multicolumn{6}{|c|}{Max-pool} \\
    \hline
    Conv5-64 & Conv3-bn-128 & Conv3-bn-256 & Conv3-bn-512 & Conv3-bn-128 & Conv3-bn-256 \\
      & Conv3-bn-128 & Conv3-bn-256 & Conv3-bn-512 & Conv3-bn-128 & Conv3-bn-256 \\
      &  &  &  & Conv3-bn-128 &  \\
      &  &  &  & Conv3-bn-128 &  \\
    \hline
    \multicolumn{6}{|c|}{Max-pool} \\
    \hline
    Fc-64 & Conv3-bn-256 & Conv3-bn-512 & Conv3-bn-1024 & Conv3-bn-256 & Conv3-bn-512 \\
      &  &  &  & Conv3-bn-256 & Conv3-bn-512 \\
    \cline{2-6}
    & \multicolumn{5}{|c|}{Max-pool} \\
    \cline{2-6}
      &  &  &  &  & Conv3-bn-512 \\
      &  &  &  &  & Conv3-bn-512 \\
    \cline{6-6}
      &  &  &  &  & Max-pool \\
    \hline
    Fc-10 & Fc-10 & Fc-10(100) & Fc-100 & Fc-10 & Fc-10(100) \\
    \hline
  \end{tabular}
  \caption{Network configurations (shown in columns).  The
convolutional layer parameters are denoted as ``Conv$\langle$receptive field size$\rangle$-bn-$\langle$number of channels$\rangle$''.  The
Fully connected layer parameters are denoted as ``Fc-$\langle$number of units$\rangle$''.} \label{fnet}
\end{table}

	Generally speaking, our results further confirm the non-uniqueness phenomenon of object representation under the goal-driven approach. We systematically investigated the representation differences between a similar performing pair of DCNNs on the two public object image datasets, CIFAR-10 that contains 60,000 images belonging to 10 categories of objects and CIFAR-100 that contains 60,000 images belonging to  100 categories of objects \cite{Krizhevsky2009}. 
In our experiments, 5,000 images per category in CIFAR-10 (also 500 images per category in CIFAR-100) were randomly selected for network training, and the rest for testing. Six network  architectures with different configurations (denoted as $\{$D1, D2, D3, D4, D5, D6$\}$) were employed for evaluations, where $\{$D1, D2, D3, D5, D6$\}$ were for CIFAR-10 and  $\{$D3, D4, D6$\}$ were for CIFAR-100 as shown in Table \ref{fnet}.

 The traditionally used measure, ``explained variance''(EV), was employed to access the degree of linearity between the learnt object representations from a similar performing pair of DCNNs, and we trained similar performing pairs of DCNNs  under the following two schemes:
\begin{itemize}
\item[Scheme-1]  Both DCNN$_1$ and
	DCNN$_2$ were trained with random initializations.
\item[Scheme-2]  Similar to the training procedure in the  DCNN$_1$ was firstly trained with the Softmax loss, and then DCNN$_2$ was trained by combining the Softmax loss on the neuron outputs of the last layer and the Euclidean loss on the differences between the neuron outputs of the penultimate layer in DCNN$_2$ and the corresponding terms calculated according to Eq. (\ref{eq3}) (In our experiments, $f(x) = |x|\sqrt{x}$).
\end{itemize}

	Here are some main results from our experiments:

\begin{figure}[t]
\centering
  \subfigure[] {  \includegraphics[width=5.2 cm]{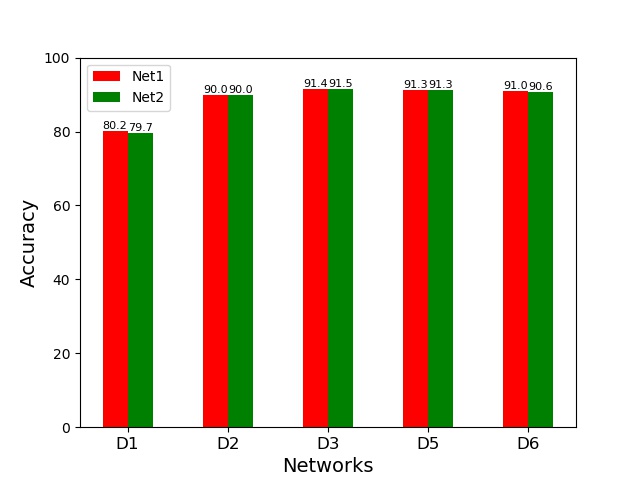} \label{f10Acc}  }
  \subfigure[] {  \includegraphics[width=5.2 cm]{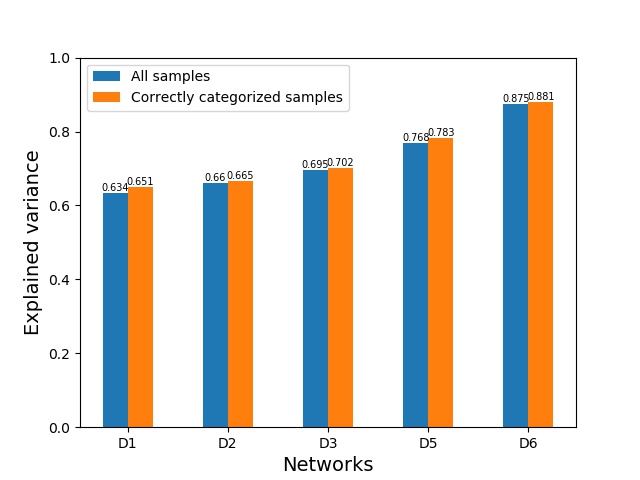} \label{f10Ev} }
  \subfigure[] {  \includegraphics[width=5.2 cm]{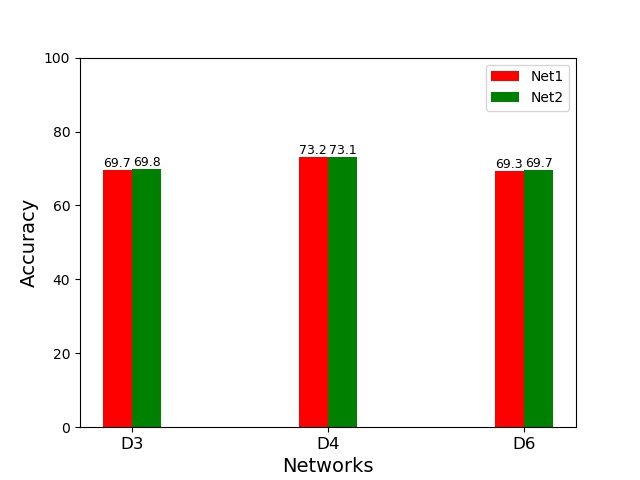} \label{f100Acc} }
  \subfigure[] {  \includegraphics[width=5.2 cm]{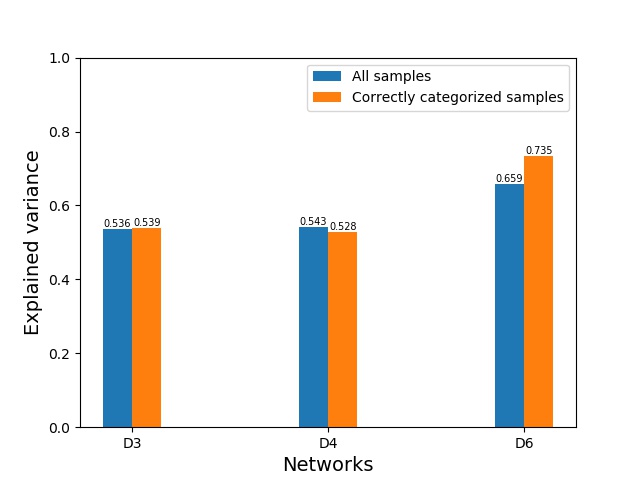}\label{f100Ev}  }
\caption{
(a) Categorization accuracies of $\{$D1, D2, D3, D5, D6$\}$ with two random initializations on CIFAR-10 (Net1 and Net2 indicate a same network with two initializations, similarly hereinafter); (b) Mean EVs on CIFAR-10 for all the inputs (blue bars)/only the  correctly categorized inputs (orange bars);(c) Categorization accuracies of $\{$D3, D4, D6$\}$ with two initializations on CIFAR-100; (d) Mean EVs on CIFAR-100 for all the inputs (blue bars)/only the  correctly categorized inputs (orange bars).} \label{fstan}
\end{figure}

\textbf{(i) Explained variance on standard data} \\
The results using the training Scheme-1 are shown in Figure \ref{fstan}.
Figure \ref{f10Acc} and Figure \ref{f100Acc} show the categorization accuracies of  similar performing pairs of DCNNs under different network architectures with two random initializations on CIFAR-10 and CIFAR-100 respectively. The blue bars of Figure \ref{f10Ev} and Figure \ref{f100Ev} show the corresponding mean EVs on CIFAR-10 and CIFAR-100 respectively.
As seen from Figure \ref{f10Ev} and Figure \ref{f100Ev}, the mean EVs by  $\{$D1, D2, D3, D5, D6$\}$ are around $63.4\% \sim 87.5\%$ on CIFAR-10, while the mean EVs by  $\{$D3, D4, D6$\}$  are around $53.6\% \sim 65.9\%$ on CIFAR-100. In addition, the mean EV of the network D1 under the training Scheme-2 is $51.2\%$ on CIFAR-10.

Two points are revealed from these results:
\begin{itemize}
\item Given a similar performing pair of DCNNs, although the representations of the two DCNNs cannot in theory be related by a linear transformation, the explained variance between the two representations is relatively large.
\item A similar performing pair of DCNNs  with a deeper architecture, or having more layers, will generally have a larger explained variance between the two representations. The underlying reason seems that since a DCNN with a deeper architecture will generally have a larger representational capacity and since a fixed task has a fixed representation demand, a DCNN with a larger capacity will give a more linear representation.
\end{itemize}

In addition, for a similar performing pair, although their categorization performances  are similar, it does not mean that the two DCNNs have the identical categorization label for each input sample, either correct or wrong. We have manually checked the categorization results for CIFAR-10 and CIFAR-100.
The orange bars of Figure \ref{f10Ev} and Figure \ref{f100Ev} show the computed mean EVs for only those inputs correctly categorized. As seen from Figure \ref{fstan}, the discrepancy of the explained variances between the representations of only the correctly categorized inputs and those of the whole inputs is insignificant and negligible in most cases, and it is perhaps due to the already high categorization rate of the two DCNNs such that the incorrectly categorized inputs only take a small fraction of a relatively large test set.

\begin{figure}[t]
\centering
  \subfigure[] {  \includegraphics[width=5.5 cm]{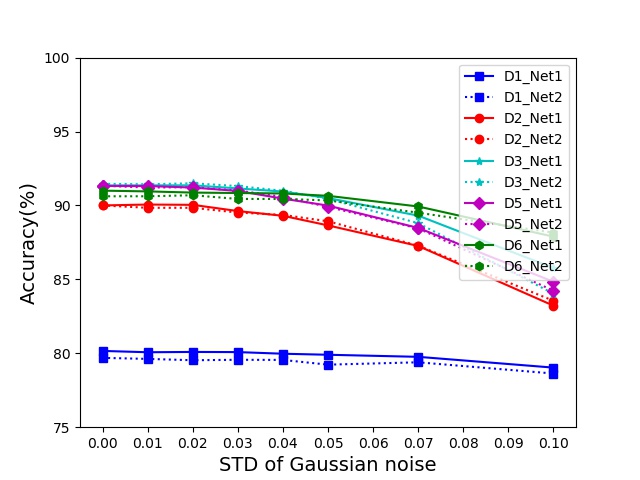} \label{fnoiseAcc} }
  \subfigure[] {  \includegraphics[width=5.5 cm]{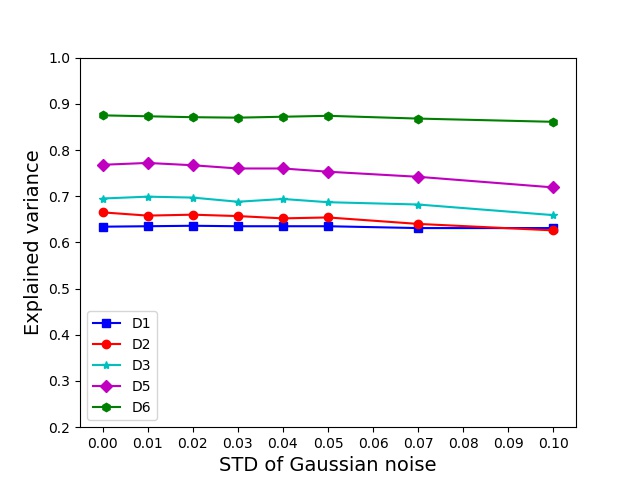}  \label{fnoiseEV} }
\caption{Categorization accuracies and mean EVs under different levels of noise: (a) Categorization accuracies of similar performing pairs of DCNNs; (b) Mean EVs  of similar performing pairs of DCNNs.}
\end{figure}

\textbf{(ii) Explained variance on noisy data} \\
In \cite{Szegedy2014}, it is reported that DCNNs are sometimes sensitive to adversarial images, that is, images slightly corrupted with random noise, which do not pose any significant problem for human perception, but dramatically alter the categorization performance of DCNNs. Here, we assessed the noise effects on the representation equivalence on CIFAR-10. The input images are normalized to the range $[0, 1]$, and Gaussian noise with mean 0 and standard variance $\sigma = \{0.01,0.02,0.03,0.04, 0.05, 0.07, 0.1\}$ are added into these images respectively.
Figure \ref{fnoiseAcc} shows the corresponding categorization accuracies of similar performing pairs of DCNNs under different architectures, while Figure \ref{fnoiseEV} shows the corresponding mean EVs. We find that even under the noise level $\sigma  = 0.1$,   the explained variance does not change much, although the categorization accuracy decreases notably.

\begin{figure}[t]
\centering
  \subfigure[] {  \includegraphics[width=5.5 cm]{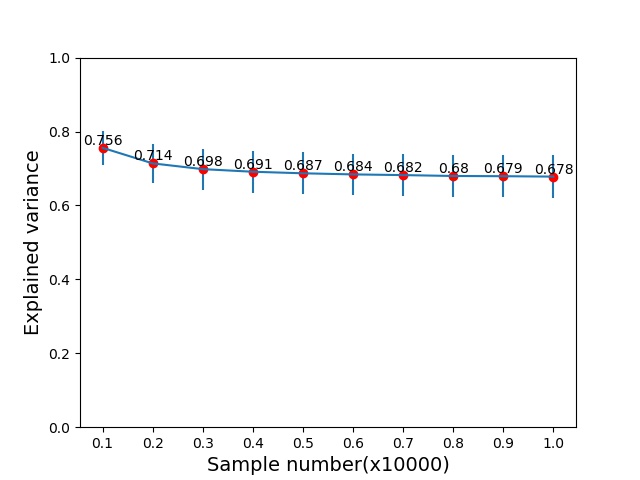} \label{fsizeall} }
  \subfigure[] {  \includegraphics[width=5.5 cm]{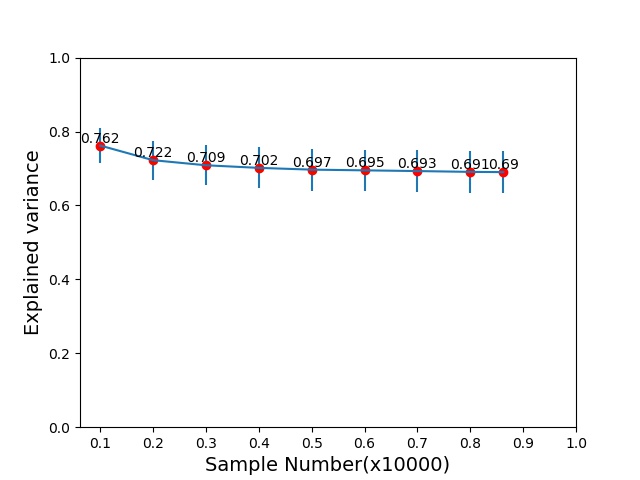}  \label{fsizecorr} }
\caption{Mean EVs with different image samples: (a) Samples are randomly selected from the whole test image set; (b) Samples are randomly selected from the set of only those  correctly categorized images.}
\end{figure}

\textbf{(iii)  Variations of explained variance by changing stimuli size}\\
In the neuroscience, the number of stimuli could not be too large. However, for image categorization by DCNNs, the size of the test set could be very large. Does the size of stimuli set play a role on the explained variance? To address this issue, we assessed the explained variance as the dataset size increases by resampling subsets from the original test set of images in CIFAR-10. Here, image subset sizes of $[1000, 2000, \cdots, 10000]$ are evaluated. Figure \ref{fsizeall} and Figure \ref{fsizecorr} show the results on the resampled subsets from the whole set of test data and the set of only those images which are correctly categorized respectively. Our results show that if the size of the stimuli set reaches a modestly large number (around $3000$), the explained variance stabilized. That is to say, we do not need a too large number of stimuli for reliably estimating explained variance. In other words, stimuli in the order of thousands could already reveal the essence, and a further increase of stimuli could not alter much the estimation.



\begin{figure}[t]
\centering
  \subfigure {  \includegraphics[width=5.5 cm]{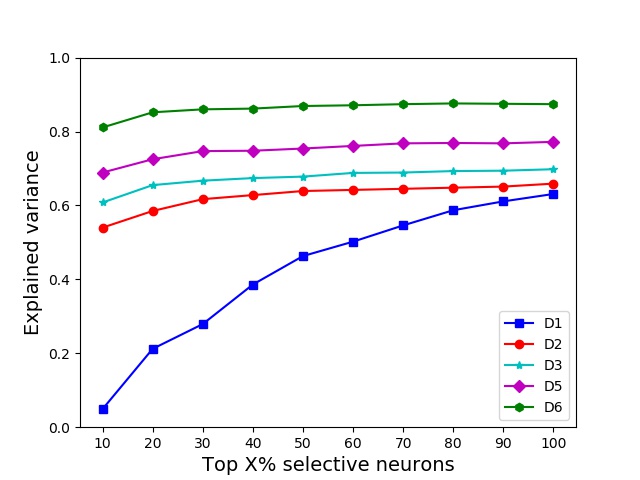} }
\caption{Mean EVs with different percentages of selective neurons.}  \label{fsele}
\end{figure}

\textbf{(iv) Explained variance vs neuron selectivity}\\
Clearly, some DCNN neurons are more selective than others \cite{Dong2017,Dong2018}. Using the kurtosis \cite{Lehky2011} of the neuron's response distribution to image stimuli, we investigated whether neuron selectivity has some correlation with the explained variance. We chose top $\{10\%, 20\%, \cdots, 100\%\}$  most selective neurons from each DCNN in a similar performing pair respectively, then computed the explained variance between the two chosen subsets, and the results are shown in Figure \ref{fsele}. As seen from Figure \ref{fsele}, with the increase of the percentage of selective neurons, the explained variance increases accordingly. This indicates that for the object representations of a similar performing pair of DCNNs, neuron selectivity is also an influential factor on their explained variance.
The explained variance between the subsets of more selective neurons is smaller, and this result seems to be in concert with the conclusion in \cite{Morcos2018}
where it is shown that neuron selectivity does not imply the importance in object
generalization ability.

\textbf{(v) A good representation does not necessarily needs IT-like}\\
In the literature \cite{Khaligh2014}, it is shown that if an object representation is IT-like, it can give a good object recognition performance. This work shows that the inverse is not necessarily true, at least theoretically speaking. That is, as shown in the above experiments and discussions,  many different representations can give the same or quite similar recognition results with/without noise.

\noindent \textbf{Remark 6:}
In this work, we assume the final classifier is a Softmax classifier. For other linear classifiers, the general concluding remark of non-equivalence can be similarly derived. Of course, if the used classifier is a nonlinear one, or the output of the penultimate layer is further processed by a nonlinear operator before inputting it to a linear classifier, as done in \cite{Chang2017}, where a 3-order polynomial is used as a preprocessing step for the final classification, our results will no longer hold. But as shown in \cite{Majaj2015}, monkey IT neuron responses can be reliably decoded by a linear classifier, we thought using Softmax as the final classifier for DCNN-based IT cortex modelling could not constitute a major problem for our results.

\section{Conclusion}
Here, we would say that we are not against using DCNNs to model sensory cortex. In fact, its potential and usefulness have been demonstrated in \cite{Yamins2014,Yamins2016}. Here, we only provide a theoretical reminder on the possible non-uniqueness phenomenon of the learnt object representations by DCNNs, in particular, by the goal-driven approach proposed in \cite{Yamins2016}.  As shown in the convergent-learning literatures, such a non-uniqueness phenomenon is prevalent in deep learning, hence when DCNNs are used for modelling sensory cortex as a general framework, people should be aware of this potential and inherent non-uniqueness problem, and appropriate network architectures in DCNN learning should be carefully considered.

\section*{Author Contributions}
Zhanyi Hu conceived of the non-uniqueness phenomenon of object representation in
modelling IT cortex by DCNN. Qiulei Dong and Zhanyi Hu explored the method. Qiulei
Dong and Bo Liu implemented the explored method and performed the validation.
Qiulei Dong and Zhanyi Hu wrote the paper.

\section*{Acknowledgements}
This work was supported by the Strategic Priority Research Program of
the Chinese Academy of Sciences (XDB32070100), and National Natural Science
Foundation of China (U1805264, 61573359).

\section*{Appendix}

\textbf{Procedure to train DCNN$_1$ and DCNN$_2$:}

\textbf{Input:}   A set of $n$ image objects: $D=\{I_i,i=1,2,\cdots,n\}$ with known categorization labels.

\textbf{Output:}  DCNN$_1$ and DCNN$_2$ whose object representations are different but with the same (or similar) categorization performance;

\begin{itemize}
\item[1] Using $D=\{I_i, i = 1, 2, \cdots, n\}$ to train a DCNN by optimizing the categorization performance. This training can be done similarly as reported in numerous image categorization literatures. Denote the trained
DCNN as DCNN$_1$. The output of the penultimate layer in DCNN$_1$ for $D$ is denoted as
$X = \{x_i, i = 1, 2, \cdots, n\}$, $x_i$ is the output for input image $I_i$. Denote the output of the final layer in DCNN$_1$ for $D$ as: $X^{'}=\{x_i^{'},i=1,2,\cdots, n\}$, the weighting matrix at the final layer in DCNN$_1$ is $W_1$ and the bias vector is $b_1$, that is $x_i^{'}= W_1 x_i + b_1$;
\item[2] Choose a nonlinear monotonically increasing function $f(\cdot)$, and compute $Y^{'}=\{y_i^{'},i=1,2,\cdots,n\}$, where $y_i^{'}=f(x_i^{'})$ in element-wise mapping;
\item[3] Choose a weighting matrix  $W_2$ for the second DCNN, say $W_2=W_1$;
\item[4] Compute $Y=\{y_i,i=1,2,\cdots,n\}$   by $y_i =(W_2^T W_2)^{+} W_2^T (y_i^{'} - b_2)$;
\item[5] Using training pair $\{(I_i \leftrightarrow y_i ),i=1,2,\cdots,n\}$ to train the second DCNN to minimize the Euclidean loss between the DCNN's output $\tilde{y}_i$  and $y_i$.
\item[6] The trained DCNN in step (5) is our required DCNN$_2$. The object representation  $x_i$ of DCNN$_1$ and $y_i$ of DCNN$_2$ are different representations by Definition \ref{def2}, because for the same object $I_i$, $x_i$ and $y_i$ can give the same categorization results in theory without noise, or similar results with noise in practice, but they cannot be transformed by a linear transformation as shown in Proposition 2.
\end{itemize}

\end{document}